\documentclass[aps,prb,linenumbers,superscriptaddress, 12pt,amsmath,amssymb,longbibliography,
preprint,
]{revtex4-2}
\newcommand{\figurewidth}{.49\columnwidth}

\usepackage{graphicx}
\usepackage{siunitx}	
\usepackage{amssymb}
\usepackage{datetime}
\usepackage{booktabs}
\usepackage{hyperref}
\usepackage[noabbrev, capitalise, nameinlink]{cleveref}
\renewcommand{\autoref}[1]{\cref{#1}}
\usepackage{xcolor}
\hypersetup{
    colorlinks,
    linkcolor={red!50!black},
    citecolor={blue!50!black},
    urlcolor={blue!80!black}
}

\usepackage{todonotes}

\newcommand{\LaH}{\texorpdfstring{La$_4$H$_{23}$}{La4H23}}
\newcommand{\LuH}{\texorpdfstring{Lu$_4$H$_{23}$}{Lu4H23}}
\newcommand{\BaH}{\texorpdfstring{Ba$_4$H$_{23}$}{Ba4H23}}
\newcommand{\EuH}{\texorpdfstring{Eu$_4$H$_{23}$}{Eu4H23}}
\newcommand{\LiNaH}{\texorpdfstring{LiNi$_3$H$_{23}$}{La4H23}}
\newcommand{\LaBH}{\texorpdfstring{LaBH$_{8}$}{LaBH8}}
\newcommand{\LaHiii}{\texorpdfstring{LaH$_3$}{LaH3}}
\newcommand{\LaHiv}{\texorpdfstring{LaH$_4$}{LaH4}}
\newcommand{\LaHix}{\texorpdfstring{LaH$_9$}{LaH9}}
\newcommand{\LaHx}{\texorpdfstring{LaH$_{10}$}{LaH10}}
\newcommand{\HS}{\texorpdfstring{H$_3$S}{H3S}}
\newcommand{\AB}{\texorpdfstring{NH$_3$BH$_3$}{NH3BH3}}					
\newcommand{\GPa}{\giga\pascal}		
\newcommand{\Tc}{\ensuremath{T_{\text c}}}
\newcommand{\Hcii}{\ensuremath{H_{\text c2}}}
\newcommand{\vF}{\ensuremath{v_{\text F}}}
\newcommand{\Rn}{\ensuremath{R_{\text n}}}
\newcommand{\Pmiiin}{\ensuremath{Pm\bar{3}n}}

\newcommand{\dd}{\text{d}}										

\begin{document}
\nolinenumbers

\title{High-temperature superconductivity in A15-type \LaH\ below \SI{100}{\GPa}}

\author{Sam Cross}
\email{sc16434@bristol.ac.uk}
\affiliation{HH Wills Laboratory, University of Bristol, Bristol, BS8 1TL, UK}

\author{Jonathan Buhot}
\affiliation{HH Wills Laboratory, University of Bristol, Bristol, BS8 1TL, UK}

\author{Annabelle Brooks}
\affiliation{HH Wills Laboratory, University of Bristol, Bristol, BS8 1TL, UK}

\author{William Thomas}
\affiliation{HH Wills Laboratory, University of Bristol, Bristol, BS8 1TL, UK}

\author{Annette Kleppe}
\affiliation{Diamond Light Source, Chilton, Oxfordshire, OX11 0DE, UK}

\author{Oliver Lord}
\affiliation{Department of Earth Sciences, University of Bristol, Bristol BS8 1RJ, UK}

\author{Sven Friedemann}
\email{sven.friedemann@bristol.ac.uk}
\affiliation{HH Wills Laboratory, University of Bristol, Bristol, BS8 1TL, UK}

\date{\today}



\begin{abstract}
\textbf{
High-temperature superconductivity has been observed in binary hydrides such as \LaHx\ at pressures above $\sim\SI{150}{\GPa}$. Hydrogen cage structures have been identified as a common motif beneficial for high critical temperatures \Tc. Efforts are now focused on finding hydride high-temperature superconductors at lower pressures. 
We present evidence for high-temperature superconductivity in binary \LaH\ with A15-type structure featuring hydrogen cages at a pressure of $P=\SI{95}{\GPa}$. 
We synthesise \LaH\ from a lanthanum film capped with a palladium catalyst promoting the dissociation of hydrogen. In resistance measurements, we observe superconductivity with a  transition temperature  $\Tc\sim\SI{90}{\kelvin}$. In X-ray diffraction on the same sample, we identify the A15-type cubic body centred structure of the lanthanum sublattice. From comparison with earlier XRD and structure prediction studies, we identify this phase with \LaH. Our study reinforces the concept of hydrogen cages for high-temperature superconductivity.
}
\end{abstract}

\maketitle

\section{Introduction}
High-temperature superconductivity up to $\sim\SI{250}{\kelvin}$ has been reported and independently confirmed in \LaHx\ at pressures $\sim$ \SI{180}{\GPa} \cite{Somayazulu2019,Drozdov2019}. In \LaHx\ and other hydride high-temperature superconductors, clathrate-like cages of hydrogen atoms have been identified as a common structural motif \cite{Kong2021,Pickard2020}. 
Recent efforts have been devoted to finding high-temperature superconducting phases at lower pressures \cite{Pickard2020}. Notably, the cubic $Fm\bar{3}m$ phase of the \LaHx\ system could be stabilised at moderate pressures around \SI{110}{\GPa} in compounds synthesised from an alloy of cerium and lanthanum \cite{Chen2023}. Recent XRD studies found new structures in the binary lanthanum hydride system that feature hydrogen cages and hence are strong candidates for superconductivity \cite{Laniel2022}.

The A15-type structure (space group \Pmiiin) is expected to be favourable for superconductivity as it promotes clathrate cages of hydrogen atoms around the body-centred cubic lattice of the metal atoms as illustrated in the inset of \autoref{fig:XRD} (c). In this structure, hydrogen atoms feature bond lengths in the range $\approx\SIrange{1.1}{1.3}{\angstrom}$ favourable for high-temperature superconductivity. This bond length is much larger than in hydrogen molecules at this pressure and indicates a weaker bond compared to molecular bonding.
Through the donation of electrons from the metal atoms, the hydrogen anitbonding bands can be populated leading to stabilisation of the hydrogen cages and providing an avenue for metallisation and superconductivity \cite{FloresLivas2020}.
The weakening of the H-H bond  and the formation of cages is required for hydrogen to contribute electronic states at the Fermi energy and hence for significant electron-phonon coupling involving the hydrogen atoms required for high-temperature superconductivity. 

Hydride compounds of the A15-type structure have been searched for superconductivity in computational studies, e.g., leading to the prediction of high-temperature superconductivity in \LiNaH\ \cite{An2023}. Recently, superconductivity was indeed discovered in members of this structure type, in \LuH\  at pressures above \SI{200}{\GPa} \cite{Li2023d}. Furthermore, the structure type has also been detected in X-ray diffraction studies of \BaH, \EuH, and \LaH\ \cite{PenaAlvarez2021,Semenok2020a,Laniel2022}. Hence, there is a strong interest to probe experimentally for superconductivity in these members of the A15 structure type.

The A15-type structure has recently been detected in the lanthanum hydride binary system with single-crystal XRD by Laniel et al. at pressures between \SIrange{96}{150}{\GPa} \cite{Laniel2022}. Laniel et al. synthesised \LaH\ by laser heating bulk lanthanum with paraffin oil as a hydrogen donor material. During the preparation of this manuscript, Guo et al. reported synthesis of \LaH\ from bulk \LaHiii\ and ammonia borane (\AB). Furthermore, Guo et al. detected superconductivity in their samples of \LaH\ in resistance measurements. Here, we present synthesis of \LaH\ from a thin film of lanthanum capped with palladium and using \AB\ as a hydrogen donor \cite{Buhot2020,Snider2021}. Upon laser heating, molecular hydrogen is dissociated by the palladium and diffuses into the lanthanum forming \LaH. We detect the formation of \LaH\ with XRD measurements and superconductivity through four-point resistance measurements including the characteristic suppression in magnetic fields up to \SI{12}{\tesla}.

\section{Synthesis of \LaH}

\begin{figure}
    \centering
    \includegraphics[width=\figurewidth]{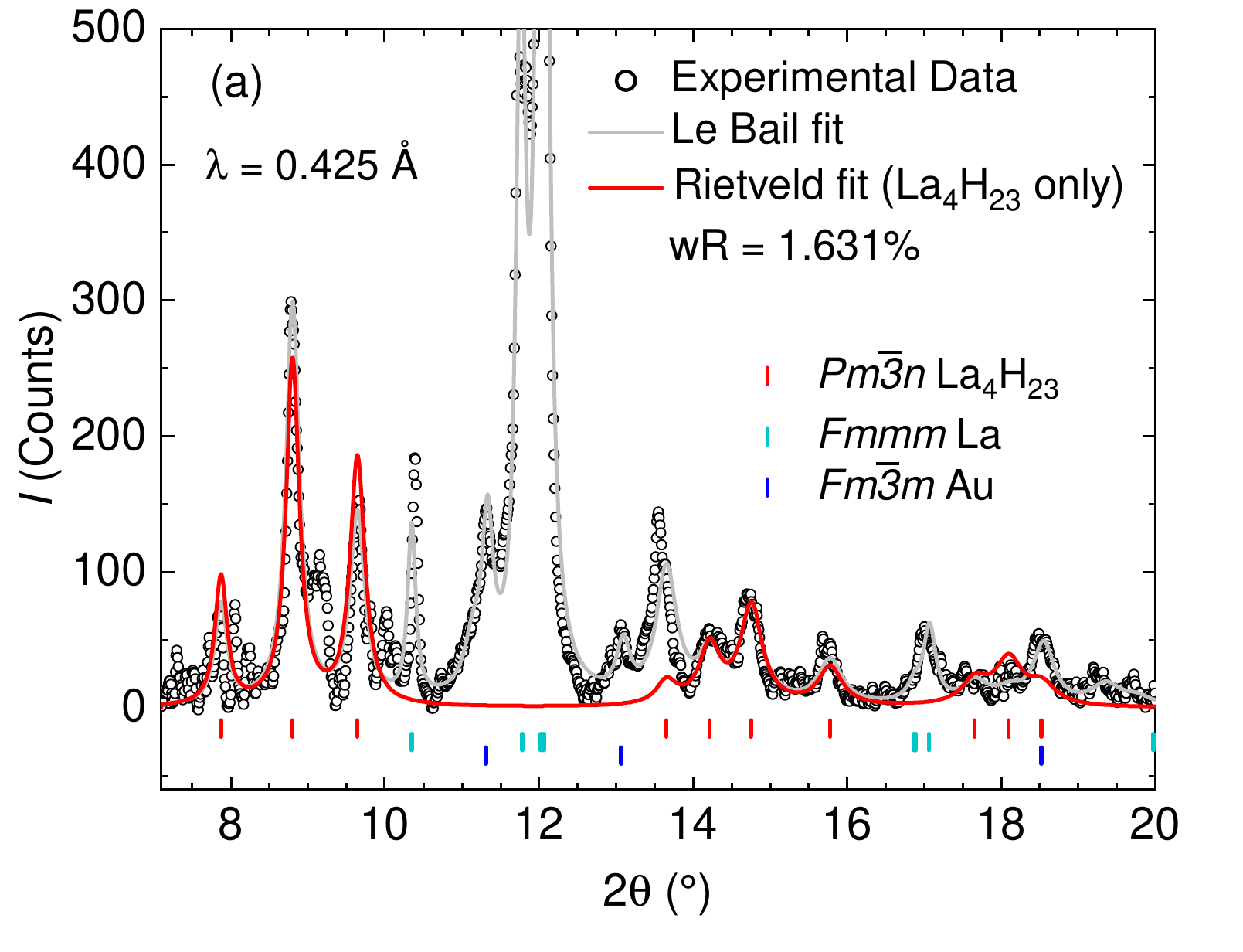}
    \includegraphics[width=\figurewidth]{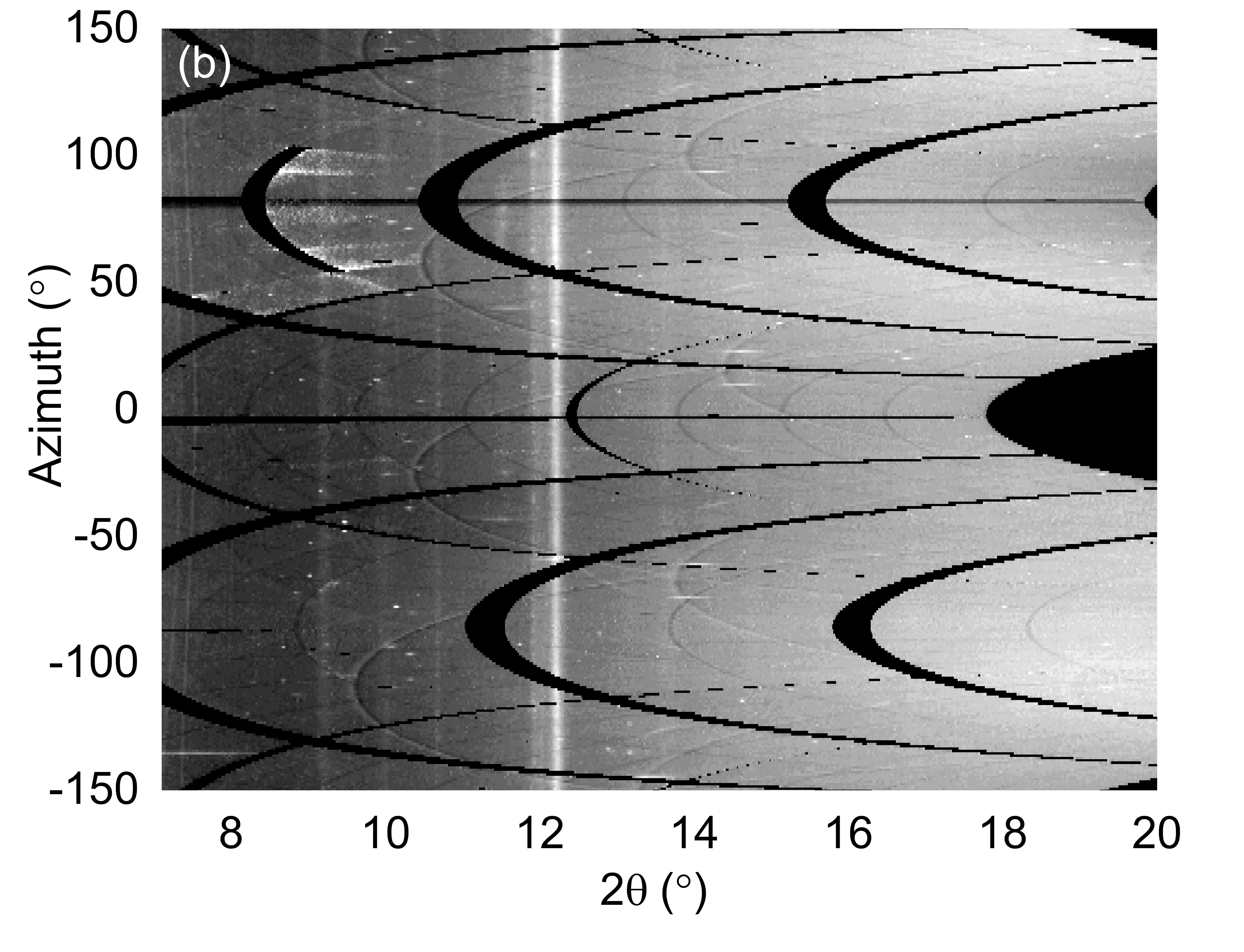}
    \includegraphics[width=\figurewidth]{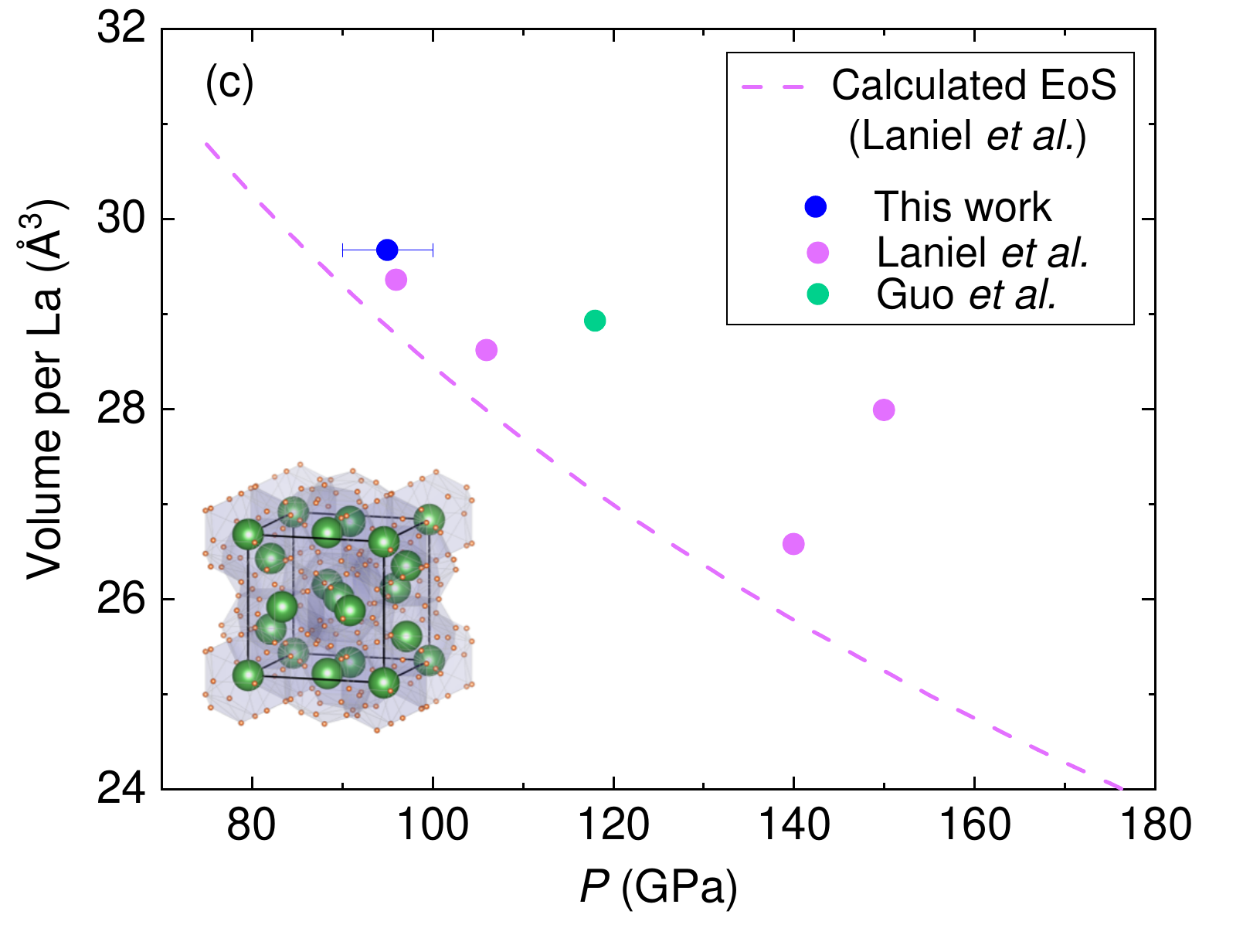}
    \caption{Detection of the A-15 structure in X-Ray diffraction. (a) Rietveld and Le Bail refinements of \LaH\ to the observed XRD pattern shown by the solid red and grey lines respectively. Data between \SIrange{11}{13}{\degree} are dominated by residual elemental lanthanum in the orthorombically distorted $Fmmm$ phase and gold in the $Fm\bar{3}m$ phase originating from the electrodes. No \LaH\ reflections are expected in this range. (b) Azimuthal (cake) plot representation of the area detector image in grayscale, where high and low intensity are represented in white and black respectively. Sample reflections appear as white vertical lines.
    (c) Unit cell volume as a function of pressure for the \LaH\ phase including data from this work, and that of references \cite{Laniel2022, Guo2023}. The A15 clathrate structure is shown in the inset, generated using the software Vesta \cite{momma2008vesta}.}
    \label{fig:XRD}
\end{figure}

The successful synthesis of \LaH\ is evident from our X-ray diffraction in \autoref{fig:XRD}. 
We observe characteristic reflections between $\SI{7.5}{\degree}\leq2\theta\leq\SI{10}{\degree}$ and $\SI{13.5}{\degree}\leq2\theta\leq\SI{16}{\degree}$ that are best assigned to the \LaH\ phase. Besides the \LaH, we also observe peaks from elemental lanthanum (orthorombically distorted $Fmmm$) and gold ($Fm\bar{3}m$). The former is residue from the synthesis of \LaH\ whilst the latter originates from the electrodes used for our resistance measurements. We can rule out several other phases including $I4/mmm$ \LaHiv\ (\autoref{fig:LaH4}), $P6_3/mmc$ \LaHix\ (\autoref{fig:LaH9}) and $C2/m$ LaH$_x$ (\autoref{fig:LaH10}) that have been reported for the La-H binary system as detailed in the extended data. We can also rule out the formation of $Fm\bar{3}m$ \LaBH\ (\autoref{fig:LaBH8}) as well as the carbohydride impurity $P6_3/mmc$ LaCH$_2$ reported by Laniel et al. at \SI{96}{\GPa} (\autoref{fig:LaCH2}) \cite{Laniel2022}.

The XRD data reveal a homogeneous sample without texture. This is visible from both the azimuthal representation of the detector image in \autoref{fig:XRD}(b) as well as the analysis of the angle-integrated XRD pattern. The detector image shows continuous vertical lines without significant spots indicating a large ensemble of grains in a polycrystalline sample. At the  same time, the angle-integrated XRD pattern exhibits  peak intensities as expected for a powder average  (\autoref{fig:XRD} (a)). Indeed, we are able to fit the observed XRD pattern with a Rietveld refinement without any preferential orientation, only invoking strain to reproduce the peak broadening.  

The lattice constants extracted from Rietveld refinement provide confidence in the assignment of the \LaH\ phase with A15-type La sublattice and suggest a large  hydrogen content close to the stoichiometric ratio La:H  = 1:5.75 for the \LaH\ phase.
The lattice parameters ($a=\SI{6.191(4)}{\angstrom}$, volume per La atom = \SI{29.66(8)}{\angstrom\cubed}) are in good agreement with previous reports and close to the predicted equation of state (EoS) (see \autoref{fig:XRD}(c)) \cite{Laniel2022}. The comparison with the EoS suggests a stoichiometric \LaH\ composition. However, some uncertainty remains for this indirect method of determining the hydrogen content as pointed out by Laniel et al. \cite{Laniel2022}. In particular, we cannot exclude vacancies on the hydrogen lattice and have limited resolution to detect the subtle modifications of the metal sublattice that might be associated with hydrogen vacancies and superstructures \cite{PenaAlvarez2021}.

\section{High-Temperature Superconductivity in \LaH}

\begin{figure}
    \centering
    \includegraphics[width=\figurewidth]{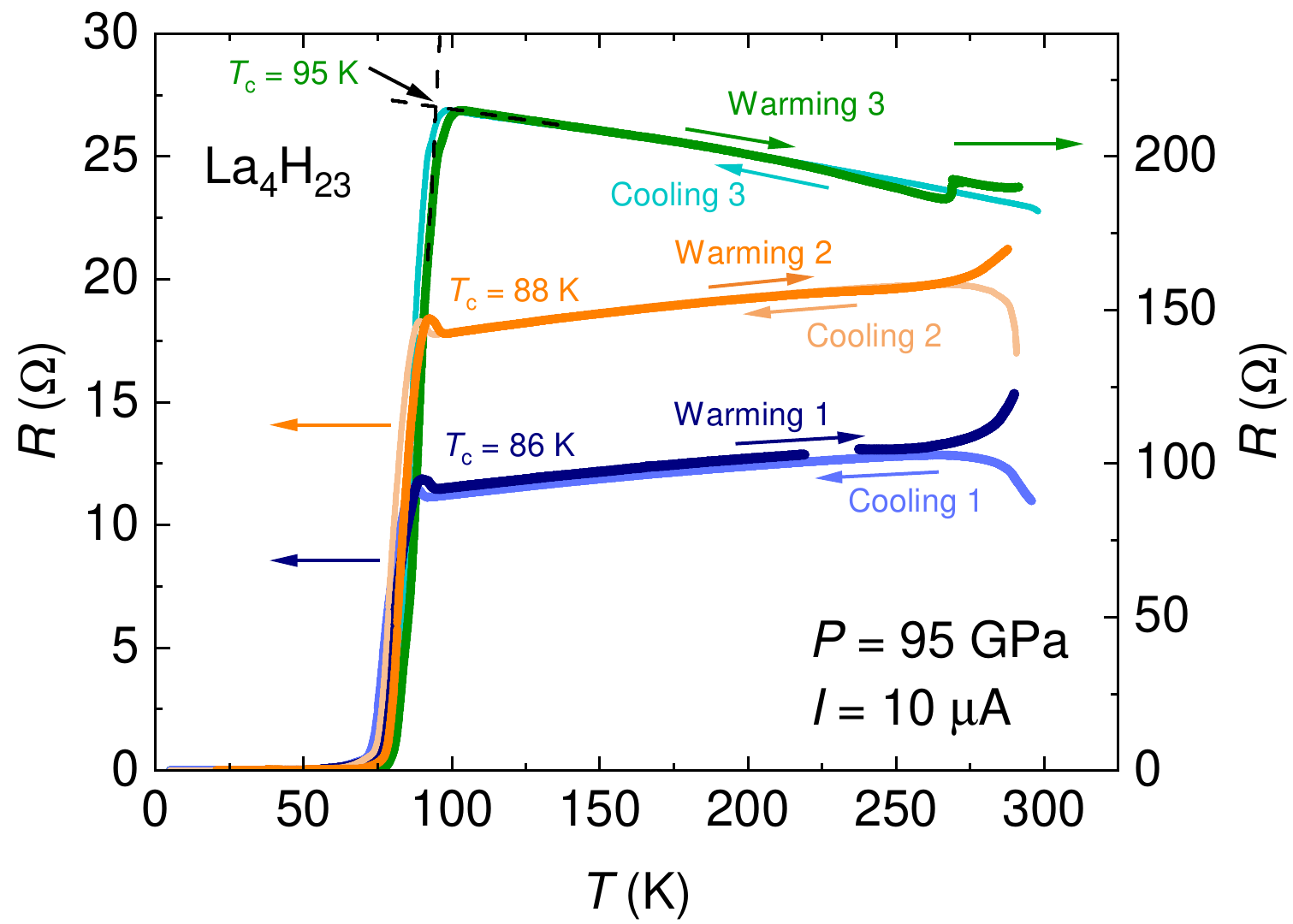}
    \caption{Detection of superconductivity in \LaH and resistance measurements on subsequent cooling runs. In each case, cooling and warming measurements are indicated. The value of \Tc\ was determined at the onset of superconductivity from the intersection of linear fits to the normal and superconducting states, shown by the dashed line in the third thermal cycle. Cooldown and warmup 1 and 2 were collected shortly after laser heating. Cooldown and warmup 3 were recorded after the XRD measurements.
    }
    \label{fig:RT}
\end{figure}

The resistance measurements presented in \autoref{fig:RT} provide evidence for high-temperature superconductivity in \LaH. Immediately after laser heating, we observe a pronounced drop in the resistance with an onset at $\Tc=\SI{86}{\kelvin}$ and reaching zero resistance within the resolution $|R-\Rn|/\Rn \leq\num{1e-3}$ of our experiment below \SI{60}{\kelvin}, where $\Rn$ is the resistance of the normal state. We have repeated resistance measurements several times including before and after XRD measurements and find the superconductivity to be stable with a slight increase to $\Tc=\SI{95}{\kelvin}$ between subsequent measurements. We attribute this increase to annealing of the sample at room temperature similar to reports on \HS\ \cite{Drozdov2015}. 

The normal state resistance shows indications of phase instabilities with decomposition exhibiting activated behaviour. If warming the sample above \SI{250}{\kelvin}, the sample resistance increases over time as illustrated in \autoref{fig:RT}. This manifests as an initial increase during cooldowns and further increase during warm-up. Only below \SI{250}{\kelvin}, cool-down and warm-up agree where we observe metallic behaviour in the  sample immediately after laser heating (thermal cycles 1 and 2).

The changes to the sample also induce fundamental change from metallic to non-metallic over time. Over multiple warm-ups to above \SI{250}{\kelvin}, the resistance changes from metallic ($\dd R/\dd T> 0$) to non-metallic ($\dd R/\dd T< 0$). A similar change from metallic to non-metallic has been observed by Guo et al. after the first laser heating and in subsequent decompression runs. In our study, the pressure before and after laser heating remained stable within our uncertainty of $\pm$ \SI{5}{\GPa}, therefore the change in the normal state could originate from either small changes in pressure and/or could be time dependent. Equally, the broadening of the superconducting transition and the loss of zero resistance reported by Guo et al. below \SI{100}{\GPa} may be related to sample decomposition.

The combination of increasing resistance and change to non-metallic behaviour suggests that part of the sample slowly transforms to a semimetallic or insulating material. In our sample, we have detected \LaH\ and elemental La with XRD, both of which may form part of the electrical path sampled by our resistance measurements with \LaH\ providing at least a percolative path supporting zero resistance of the high-\Tc\ state. Hence, the increase in resistance with time above \SI{250}{K} could be a conversion of either \LaH\ or La to a semimetallic or insulating phase, e.g. \LaHiii\ which has been reported to become semimetallic at high pressures \cite{Drozdov2019}. We argue that \LaH\ is converting because Guo et al. observe a very similar change from metallic to non-metallic but don't have elemental lanthanum in their pressure cell. We note that \LaHiii\ has been ruled out as a high-\Tc\ material \cite{Drozdov2019} and could remain undetected in XRD in the presence of \LaH\ as shown in \autoref{fig:LaHiii}. In our sample the volume fraction of \LaH\, whilst decreasing, retains a full path to detect zero resistance during the course of our measurements whilst in the study of Guo et al., the non-zero resistance below \SI{100}{\GPa} suggests that a larger portion of the sample has transformed to a semimetallic state.

The change in resistance may be related to the thermodynamic and/or dynamical instabilities found in DFT calculations by Guo et al. \cite{Guo2023}. At \SI{100}{\GPa}, the DFT predicts that \LaH\ is above the convex hull by a small energy. Such an energy scale could lead to a disintegration with an activation energy. In addition, DFT predicts imaginary phonons that indicate that the phase is inherently unstable. Such imaginary phonons are in contradiction with the observed presence of \LaH\ in the studies by Laniel, Guo, and in the present work. Hence, a mechanism like a superstructure of the hydrogen lattice must exist that stabilises \LaH\ against dynamical instabilities. This mechanism might still keep the material metastable with a low energy barrier comparable to the activation energy of $\sim\SI{250}{\kelvin}$ observed here. 

Further investigations are required to map the stability range of the \LaH\ phase including under the influence of excess or deficiency of hydrogen in the DAC. So far, all reported syntheses have employed hydrogen donor materials like \AB\ and paraffin. It will be of interest to check if a synthesis in excess molecular hydrogen shows a larger stability range to lower pressures. Further X-ray diffraction that check for superstructures and  decomposition are equally desirable but challenging.

\begin{figure}
    \centering
    \includegraphics[width=\figurewidth]{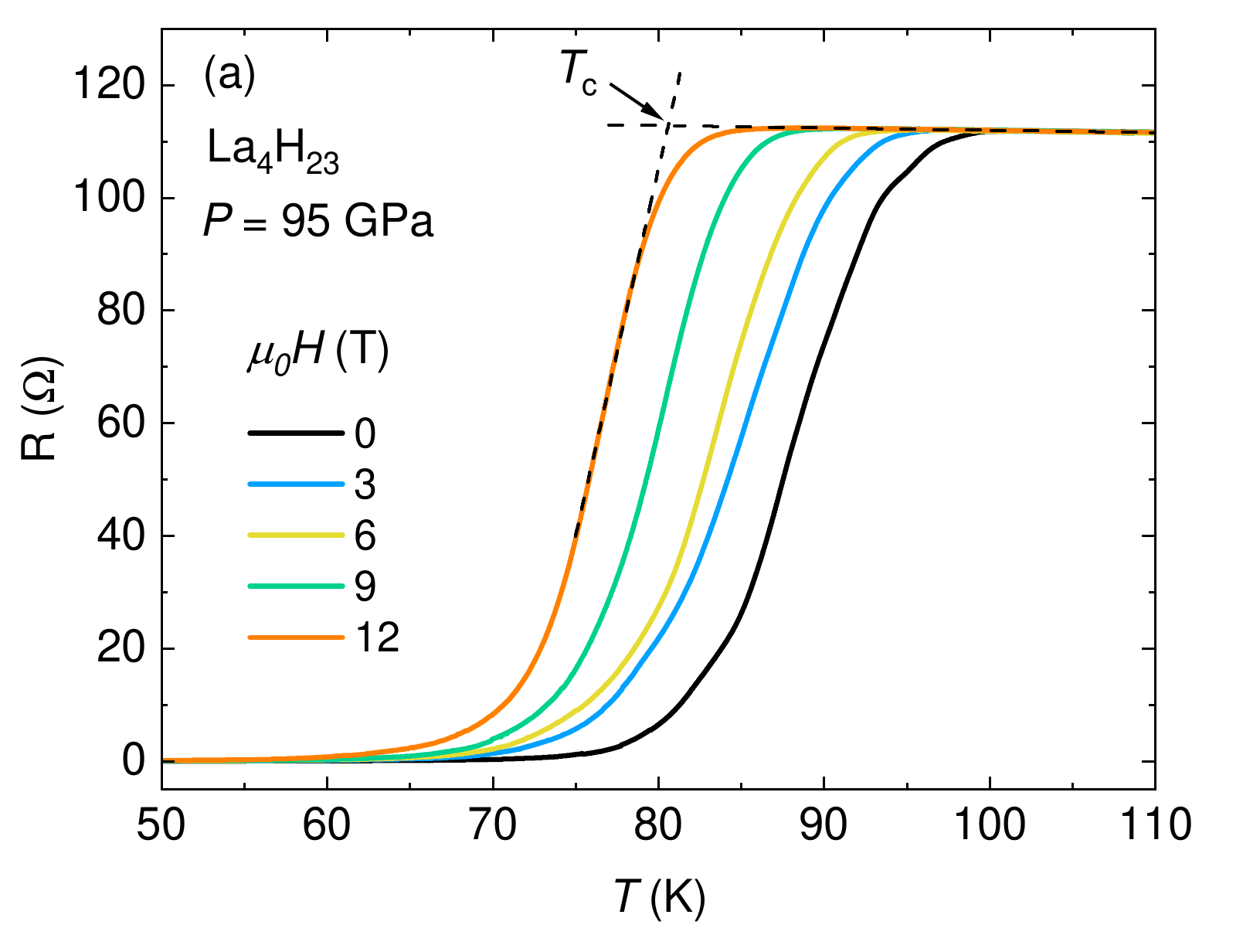}
    \includegraphics[width=\figurewidth]{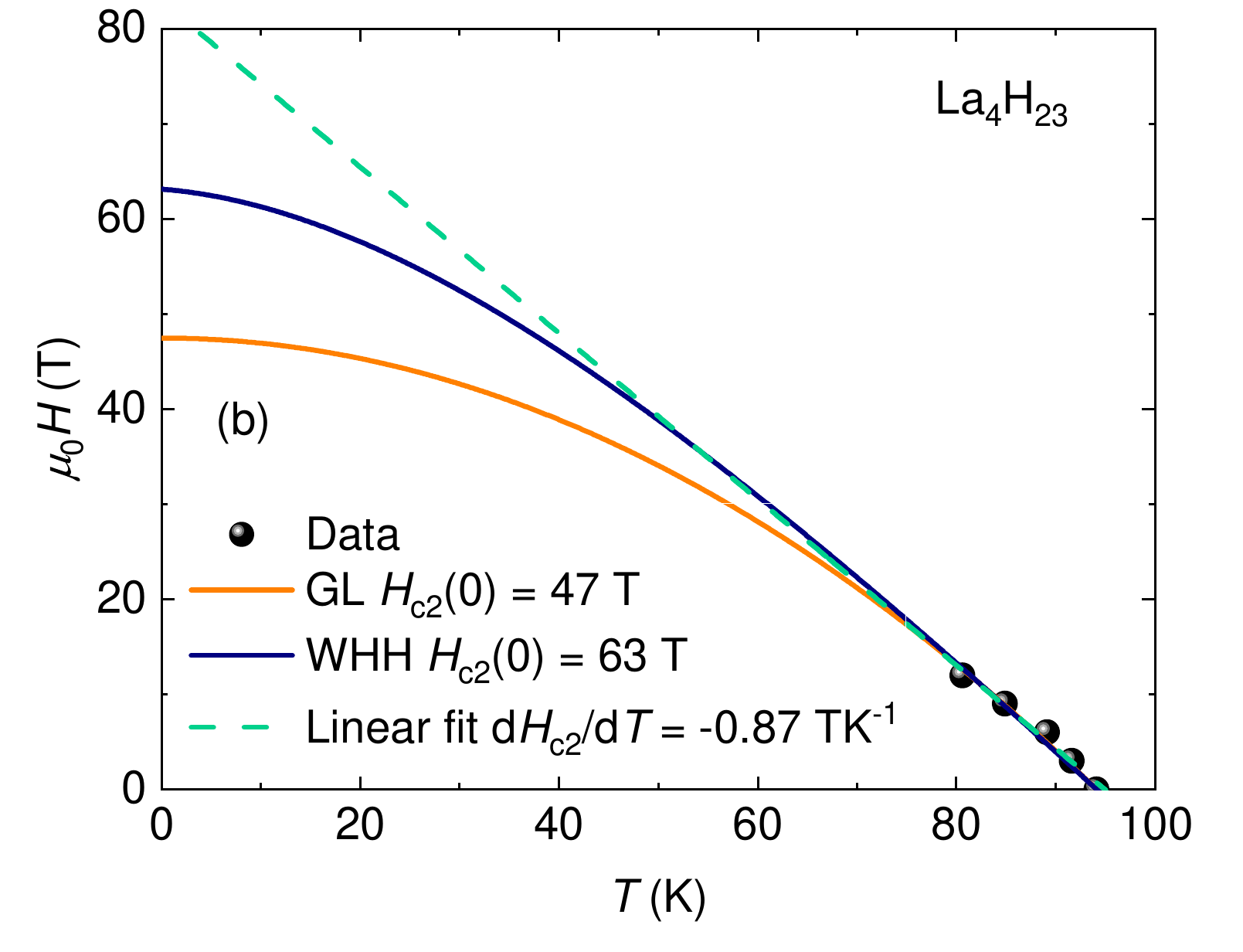}
    \caption{Superconducting critical field of \LaH. (a) Temperature sweeps for different magnetic fields. The value of \Tc\ was determined at the onset of superconductivity as shown by the dashed line for the \SI{12}{\tesla} curve. (b) Temperature dependence of the upper critical field with Ginzburg-Landau and WHH fits to estimate H$_{c2}(0)$, and a linear fit to extract the gradient near \Tc.}
    \label{fig:RB}
\end{figure}

The suppression of the resistive transition in magnetic field provides further evidence for superconductivity in \LaH. The transition is suppressed by \SI{13}{\kelvin} in a magnetic field of \SI{12}{\tesla} as illustrated in \autoref{fig:RB} (a). We extract $\Tc(H)$ in order to map the temperature dependence of the upper critical field $\Hcii(T)$ as shown in \autoref{fig:RB}(b), where the zero-temperature value $\Hcii(0)$ is estimated from extrapolations of the Ginzburg-Landau (GL) and WHH models, yielding \SI{47}{\tesla} and \SI{63}{\tesla} respectively. The linear regime of the critical field $\Hcii(T)$ can be used to estimate the zero-temperature coherence length $\xi_0$ of the superconducting state and the Fermi velocity \vF\ of the normal state. Using the clean-limit expression, and the slope of the critical field near \Tc\ ($\dd \Hcii/\dd T$ = -0.87 TK$^{-1}$) we obtain $\xi_0=\SI{2.7}{\nano\meter}$. The change of the resistance to non-metallic indicates that the sample requires a dirty-limit expression. In the absence of direct or indirect measurements that allow to estimate the mean free path, the dirty-limit expressions cannot be used to extract $\xi_0$. However, the clean-limit expression still provides a good estimate and likely only small corrections will arise from dirty-limit behaviour. Indeed the comparable slope of $\Hcii(T)$ observed by Guo et al. suggests that both studies observe the intrinsic behaviour of \LaH. 

The Fermi velocity is estimated as $\vF=\SI{1.8e5}{\meter\per\second}$. This value is lower compared to \HS\ \cite{Osmond2022}. Likely, this reflects a reduced renormalisation and hence a weaker electron-phonon coupling compared to \HS\ in agreement with the lower \Tc. A more accurate estimation of the renormalisation requires detailed band structure calculations based on a stable structure.

\section{Conclusion}
The discovery of high-temperature superconductivity in \LaH\ with the A15 structure reinforces that hydrogen cage structures are beneficial for superconductivity. The A15-type structure is so far the prime example of how high-temperature superconductivity can be realised below \SI{100}{\GPa} in a binary hydride material. 

The XRD analysis suggests that  \LaH\ is the major phase synthesised at pressures around \SI{100}{\GPa}. This may help forming pure samples in the future to study the intrinsic superconducting properties and hence has the potential to advance the understanding of high-temperature superconductivity in hydride compounds. 


\section{Methods}
High pressures were achieved in a DAC machined from MP35N alloy, with diamond anvils of culet diameter \SI{50}{\micro\metre} bevelled at \SI{8}{\degree} to a diameter of \SI{300}{\micro\metre}. Gaskets were prepared by pre-indentation of \SI{250}{\micro\metre} steel 301 and insulated with a boron nitride and epoxy mixture. The insulation was indented to a thickness $\sim\SI{30}{\micro\metre}$ and a sample space of diameter \SI{28}{\micro\metre} was drilled. The anvil with the sample was first  coated in \SI{50}{\nano\meter} of alumina to provide a thermal insulation during laser heating and to help prevent diamond embrittlement from hydrogen diffusion. 
For electrical measurements, five bi-layer tungsten-gold electrodes were sputtered/evaporated directly onto one of the diamond anvils. 

A lanthanum film of thickness \SI{260}{\nano\metre} was deposited onto the electrodes in a vacuum $\sim\SI{1e-7}{\milli\bar}$ after multiple pre-melts to remove oxide impurities from the La source. The La film was capped with a \SI{5}{\nano\meter} layer of palladium to act as an oxidation barrier, and to catalyse hydrogen diffusion into the La sample \cite{snider2021synthesis}. 

Due to the hygroscopic nature of AB, we first purified by sublimation to remove any residual moisture and impurities that would react with the La film. Loading of the AB into the gasket indentation and initial cell closing were done in an inert Ar atmosphere in a glove box with residual O$_2$/H$_2$O $<$ 0.5 ppm. The sample was pressurised to 95 GPa with \AB\ serving as the pressure transmitting medium (PTM).

The pressure was determined from the Raman shift of the diamond edge using the Akahama scale as shown in \autoref{fig:Pressure} \cite{akahama2006pressure}. The unit cell volumes extracted from Le Bail refinement of residual lanthanum (\textit{Fmmm}) and gold (\textit{Fm$\overline{3}$m}) (\autoref{fig:XRD_La}) from the electrodes are consistent with the pressure measured using the diamond edge \cite{chen2020superconductivity, takemura2008isothermal}. 

The La film was characterised in resistance measurements prior to laser heating (\autoref{fig:RT_La}), confirming metallic behaviour with residual resistance ratio ($RRR$) $\sim$ 2.5. The superconducting transition of pure La is observed with \Tc\ = \SI{8.6}{\kelvin} at \SI{95}{\GPa} which is $\sim\SI{1}{\kelvin}$ higher than reported in previous studies \cite{chen2020superconductivity}, and may originate from a pressure gradient across the sample (cf. sample pictures in \autoref{fig:sample}).

Synthesis of \LaH\ was achieved through heating of the sample and hydrogen donor material AB. The hydrogen released from AB is dissociated at the palladium capping layer and diffuses into the lanthanum film where it reacts to form \LaH. 
Heating of the sample was done using a \SI{1064}{\nano\meter} YAG laser with multiple \SI{300}{\milli\second} pulses. Laser power was incremented after each successive pulse until glowing was observed indicating a temperature of $\sim\SI{1000}{\kelvin}$. The four-point resistance was monitored during heating, where the normal state resistance increased by a factor $\sim$ 20 indicating chemical reaction between the La film and the AB. No further heating of the sample was carried out after the first observed glowing in order to preserve the electrodes, which likely explains the residual lanthanum observed in XRD.

X-ray diffraction was carried out at the I15 extreme conditions beamline at Diamond Light Source with a Pilatus3 X CdTe 2M area detector, wavelength $\lambda = \SI{0.425}{\angstrom}$ (sample-to-detector distance \SI{385}{\milli\metre}). The beam was collimated to \SI{20}{\micro\metre}. The calibrant used was LaB$_6$. Integration of the powder pattern and removal of spurious reflections were done in Dioptas \cite{prescher2015dioptas}. Background subtraction and structural refinements were done in GSAS II \cite{toby2013gsas}.

Electrical transport measurements were done in a four-point configuration with an excitation current of \SI{10}{\micro\ampere} using either an SRS SIM921 AC resistance bridge or an SR830 lock-in amplifier. Magnetic field studies were done at the University of Bristol with an Oxford Instruments \SI{14}{\tesla} magnet.

\begin{acknowledgements}

The authors are grateful for discussions with Lewis Conway. This work was carried out with the support of Diamond Light Source, instrument I15 (proposal CY35426-1). This work was supported by ERC Horizon 2020 programme under grant 715262-HPSuper and the EPSRC under grants EP/V048759/1. 
\end{acknowledgements}

\section*{Additional information}

Data will be made available at the University of Bristol data repository 

\bibliography{LaH.bib}

\newpage
\section*{Extended Data}

\renewcommand{\thefigure}{E\arabic{figure}}
\renewcommand{\thetable}{E\arabic{table}}
\renewcommand{\thesection}{S \Roman{section}}
\renewcommand{\theequation}{S\arabic{equation}}
\setcounter{figure}{0}
\setcounter{table}{0}
\setcounter{section}{0}
\setcounter{equation}{0}




\begin{figure}[h]
    \centering
    \includegraphics[width=\figurewidth]{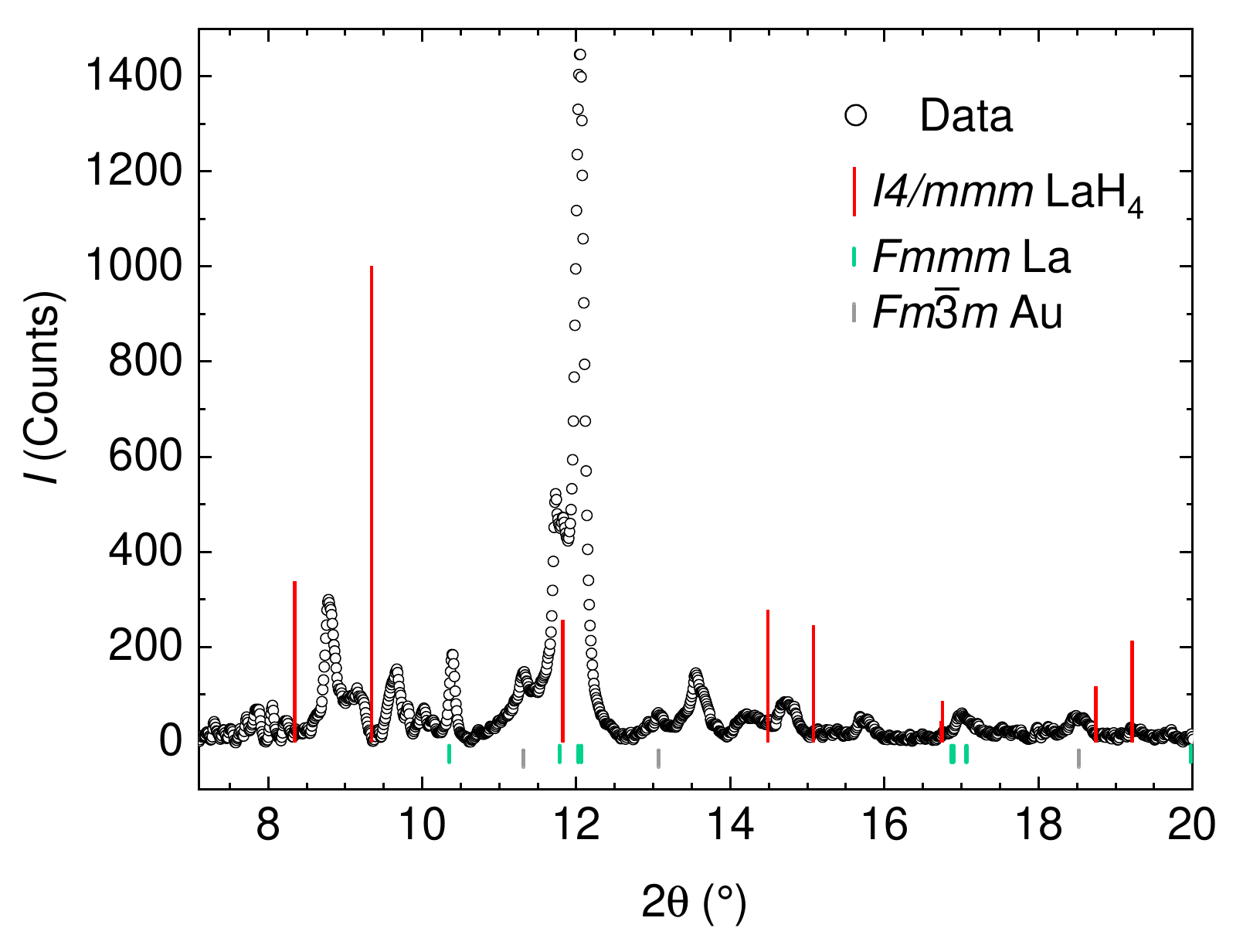}
    \caption{Peak positions and relative intensities for $I4/mmm$ LaH$_4$, generated with lattice parameters reported by Chen et al. at \SI{108}{\GPa} \cite{Chen2023}: $a = b = \SI{2.918}{\angstrom}$ and $c = \SI{5.841}{\angstrom}$. To our knowledge, there has been no experimental observation of the $I4/mmm$ phase below this pressure.}
    \label{fig:LaH4}
\end{figure}

\begin{figure}[h]
    \centering
    \includegraphics[width=\figurewidth]{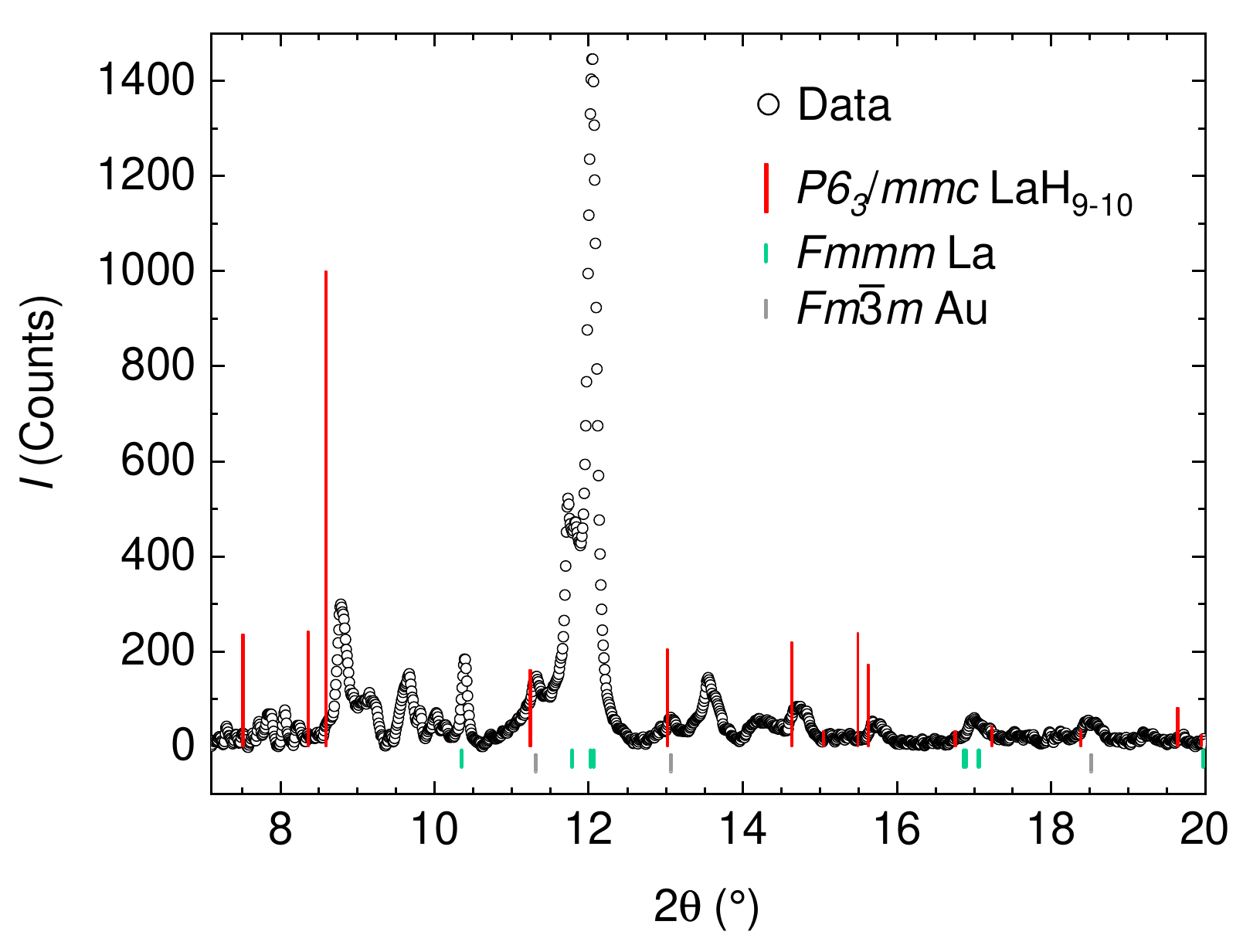}
    \caption{Peak positions and relative intensities for $P6_3/mmc$ LaH$_{9-10}$, generated with lattice parameters reported by Chen et al. at \SI{89}{\GPa} \cite{Chen2023}: $a = b = \SI{3.748}{\angstrom}$ and $c = \SI{5.833}{\angstrom}$.}
    \label{fig:LaH9}
\end{figure}

\begin{figure}[h]
    \centering
    \includegraphics[width=\figurewidth]{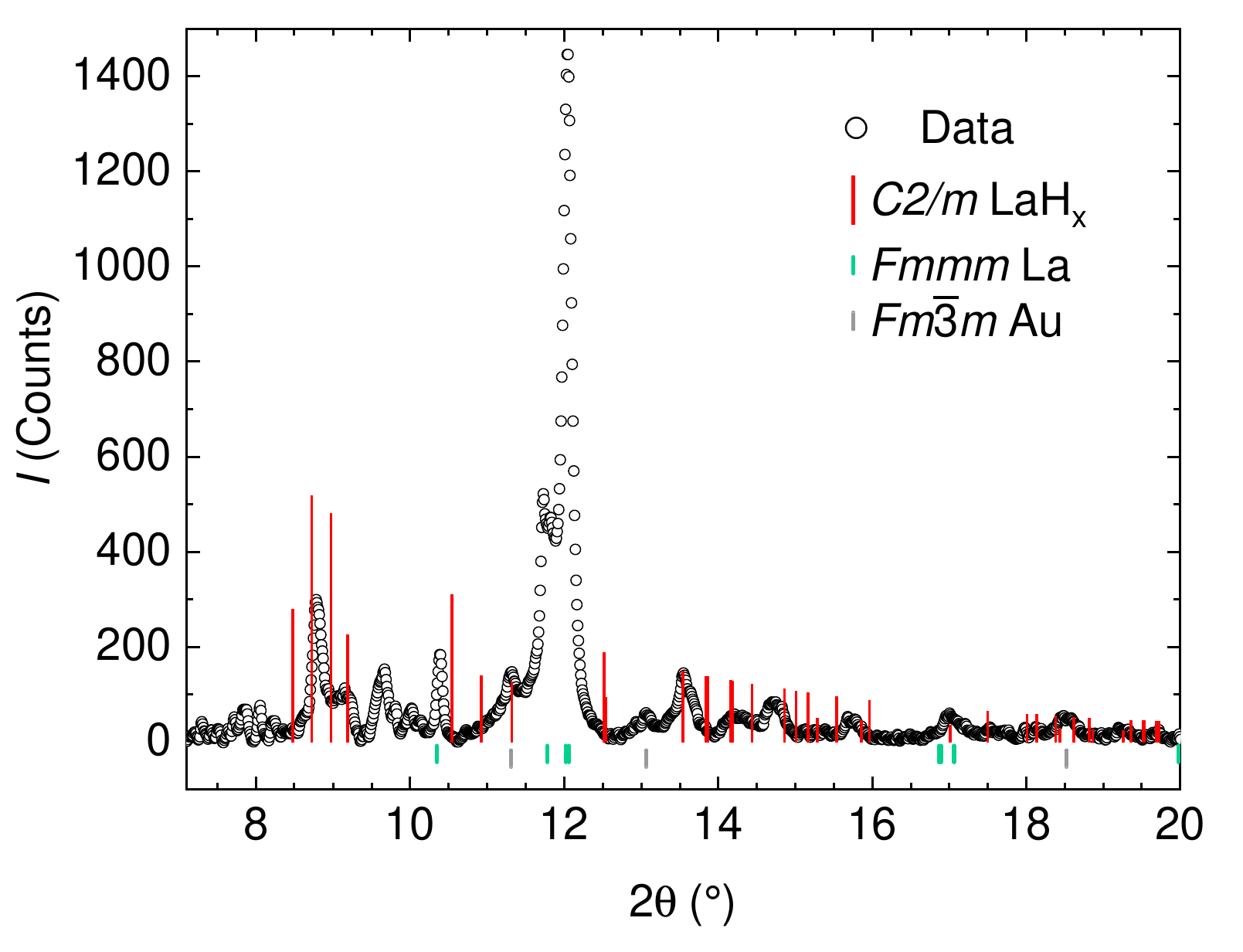}
    \caption{Peak positions and relative intensities for $C2/m$ LaH$_x$, generated with lattice parameters reported by Chen et al. at \SI{89}{\GPa} \cite{Chen2023}: $a = \SI{5.310}{\angstrom}$, $b = \SI{5.748}{\angstrom}$, $c = \SI{3.893}{\angstrom}$ and $\beta = \SI{92.2}{\degree}$.}
    \label{fig:LaH10}
\end{figure}

\begin{figure}[h]
    \centering
    \includegraphics[width=\figurewidth]{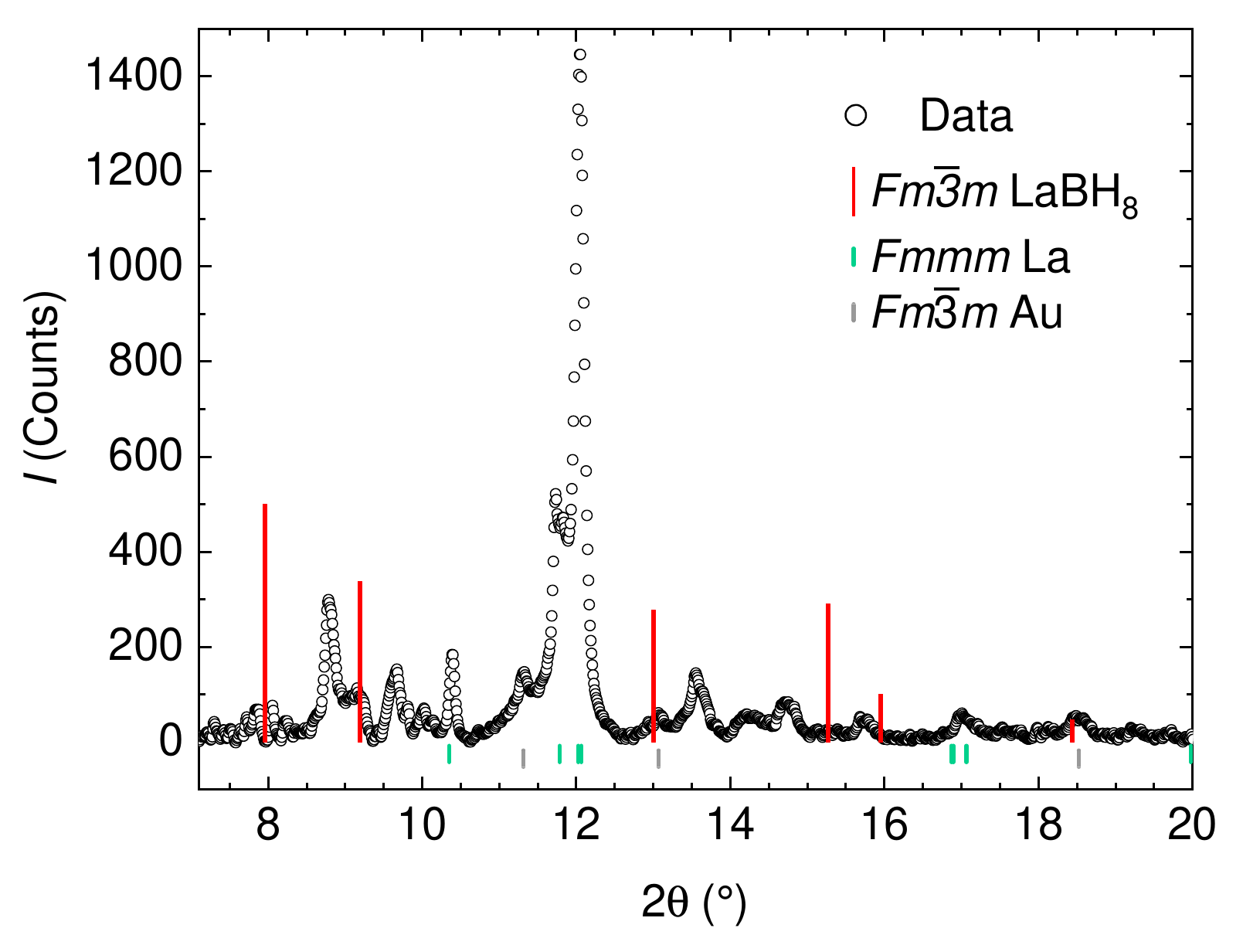}
    \caption{Peak positions and relative intensities for $Fm\bar{3}m$ LaBH$_8$, generated with lattice parameters calculated by Belli and Errea at \SI{100}{\GPa} \cite{belli2022impact}: $a = \SI{5.311}{\angstrom}$.}
    \label{fig:LaBH8}
\end{figure}

\begin{figure}[h]
    \centering
    \includegraphics[width=\figurewidth]{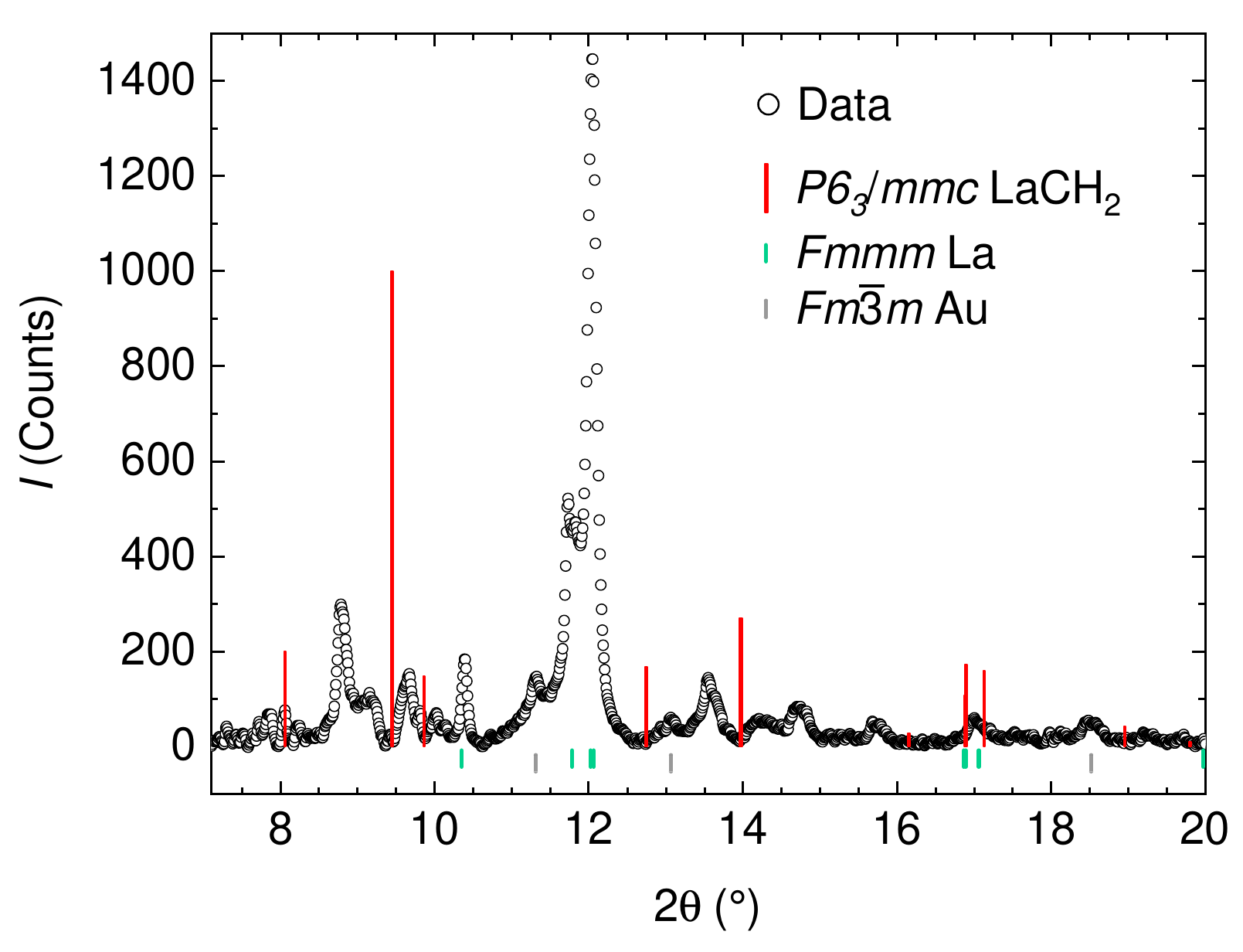}
    \caption{Peak positions and relative intensities for $P6_3/mmc$ LaCH$_{2}$ as reported by Laniel et al. at \SI{96}{\GPa}: $a=b=\SI{3.4936}{\angstrom}$, $c = \SI{4.9430}{\angstrom}$.}
    \label{fig:LaCH2}
\end{figure}

\begin{figure}[h]
    \centering
    \includegraphics[width=\figurewidth]{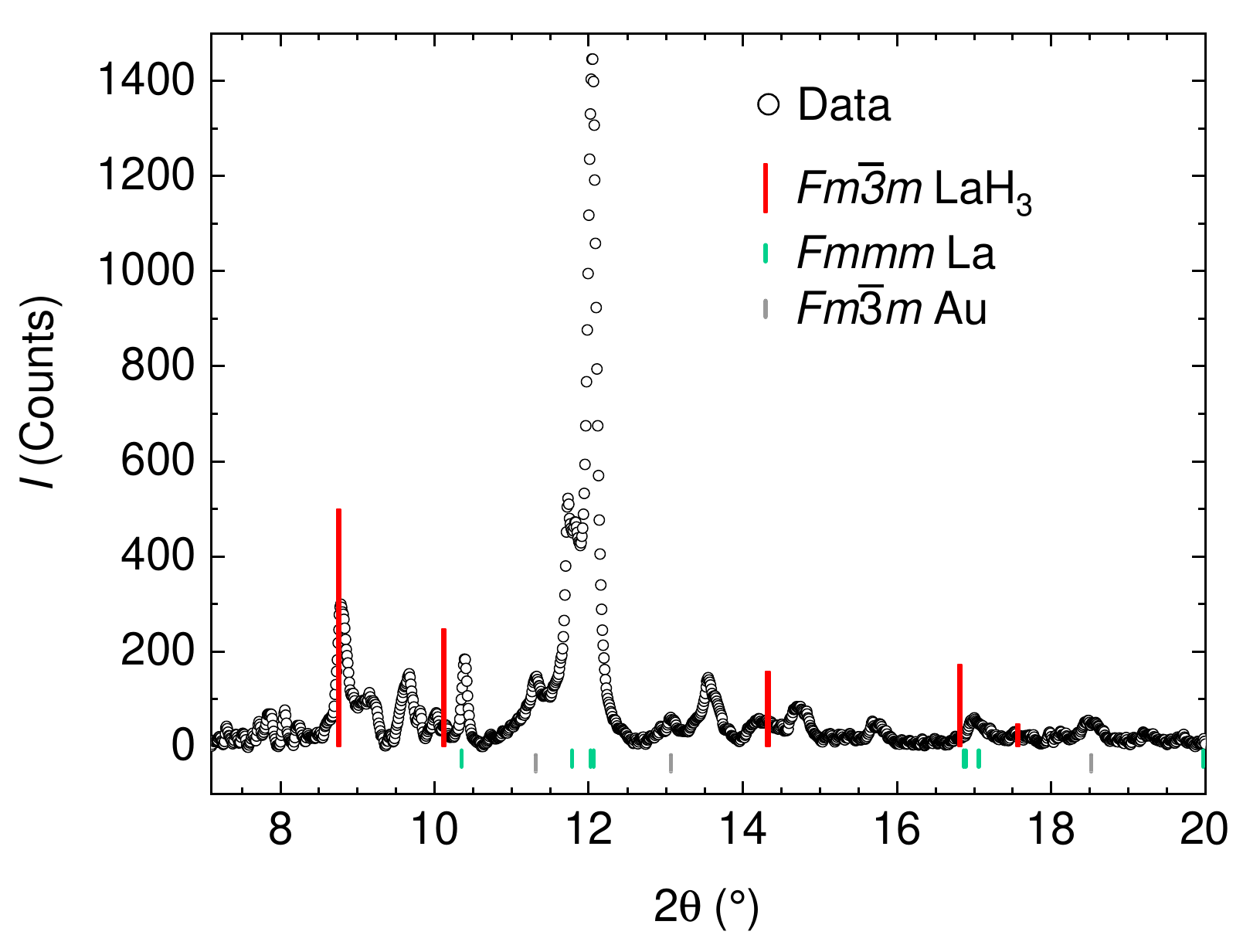}
    \caption{Peak positions and relative intensities for $Fm\bar{3}m$ LaH$_3$, generated with lattice parameter $a = \SI{4.82}{\angstrom}$ estimated from the equation of state given in reference \cite{Laniel2022}.}
    \label{fig:LaHiii}
\end{figure}

\begin{figure}[h]
    \centering
    \includegraphics[width=\figurewidth]{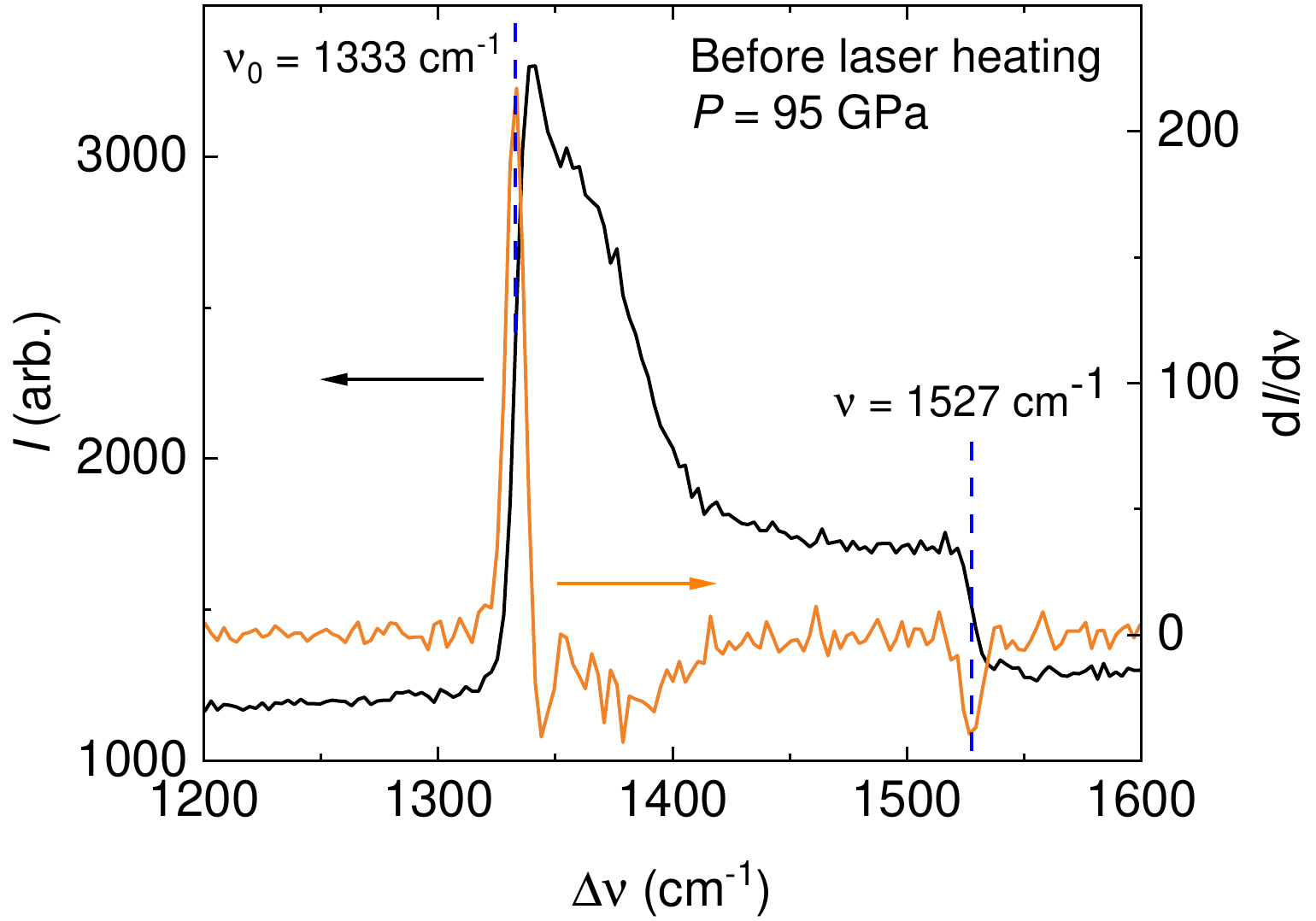}
    \caption{The determination of the pressure before laser heating from the Raman spectrum of the diamond edge using the Akahama scale \cite{akahama2006pressure}.}
    \label{fig:Pressure}
\end{figure}

\label{SI:XRD_La}
\begin{figure}[h]
    \centering
    \includegraphics[width=\figurewidth]{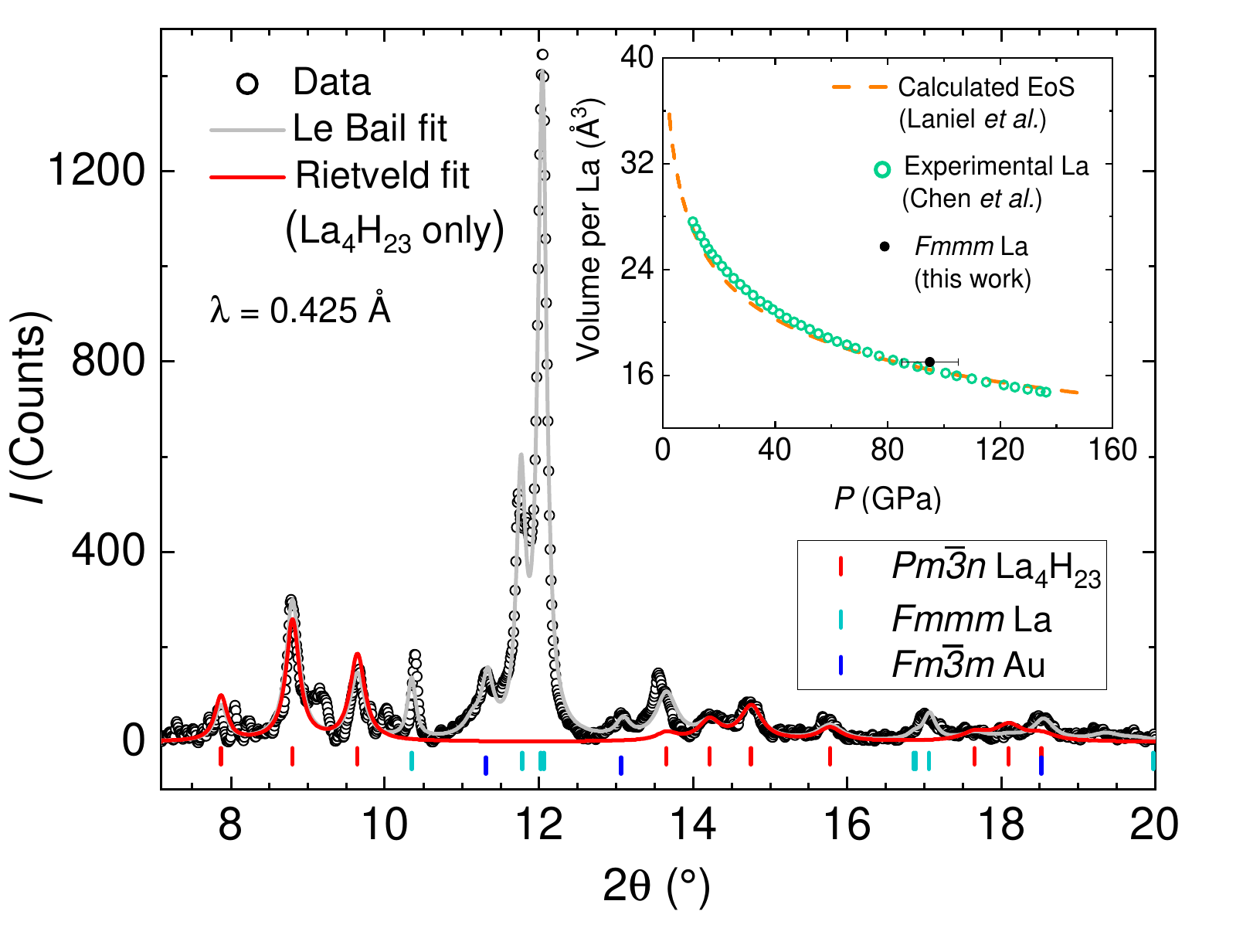}
    \caption{Le Bail and Rietveld refinement of the XRD pattern. The volume per lanthanum atom of the pure La ($Fmmm$) is \SI{16.99(9)}{\angstrom^3} in agreement with previous calculated and experimental reports, shown in the inset \cite{Laniel2022, chen2020superconductivity}.}
    \label{fig:XRD_La}
\end{figure}

\label{SI:RT_La}
\begin{figure}[h]
    \centering
    \includegraphics[width=\figurewidth]{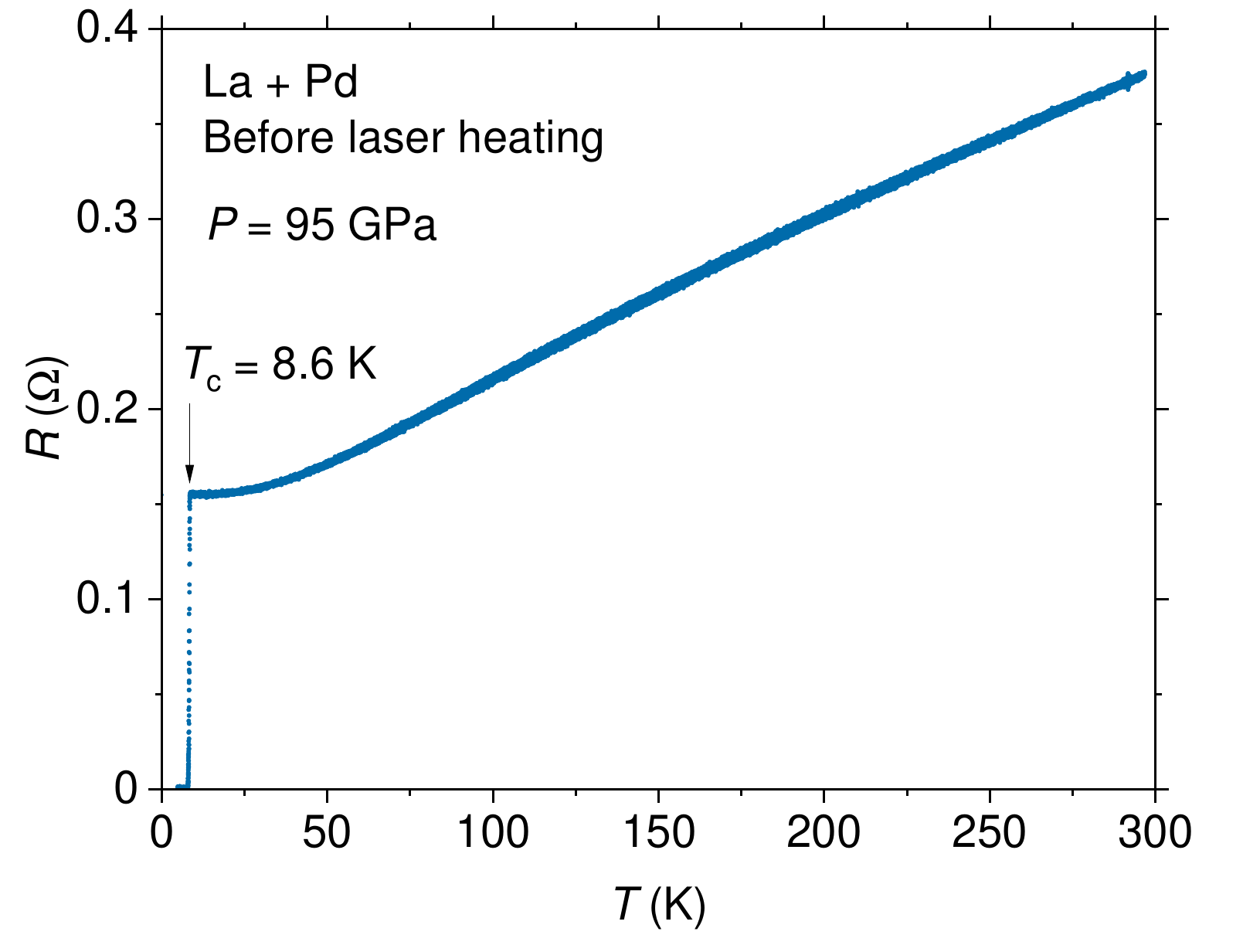}
    \caption{Resistance versus temperature of the La + Pd film at \SI{95}{\GPa} before laser heating, displaying metallic behaviour with $RRR \sim$ 2.5 and superconductivity with onset \Tc = \SI{8.6}{\kelvin}.}
    \label{fig:RT_La}
\end{figure}

\begin{figure}[h]
    \centering
    \includegraphics[width=\figurewidth]{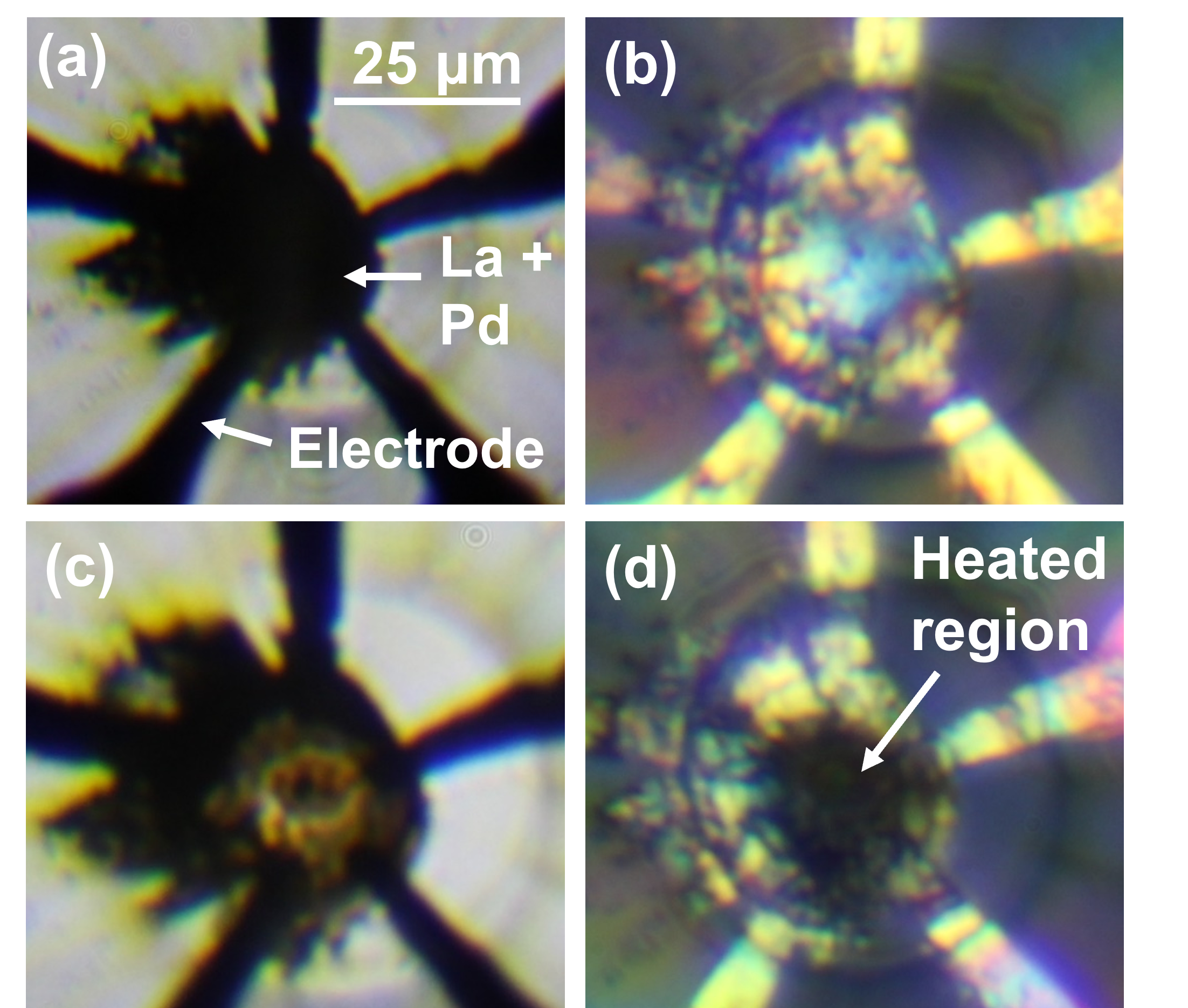}
    \caption{Sample photographs before laser heating in (a) transmission and (b) reflection. The sample after laser heating is shown in (c) and (d).}
    \label{fig:sample}
\end{figure}

\end{document}